\documentclass[12pt,preprint]{aastex}

\shorttitle{SMA OBSERVATIONS OF W3(OH) COMPLEX}

\shortauthors{Qin et al.}

\begin{document}




\title{SMA OBSERVATIONS OF W3(OH) COMPLEX: PHYSICAL AND CHEMICAL
  DIFFERENTIATION BETWEEN W3(H$_2$O) AND W3(OH) }







\author{S\sc{heng}-L\sc{i} Q\sc{in},\altaffilmark{1} P\sc{eter}
S\sc{chilke},\altaffilmark{2} J\sc{ingwen}
  W\sc{u},\altaffilmark{3} Y\sc{uefang}
W\sc{u},\altaffilmark{4} T\sc{ie} L\sc{iu},\altaffilmark{5}  Y\sc{ing}  L\sc{iu},\altaffilmark{6}  \'A\sc{lvaro}  S\sc{\'anchez}-M\sc{onge}\altaffilmark{2}}
\altaffiltext{1}{Department of
Astronomy, Yunnan University, and  Key Laboratory of Astroparticle Physics of
Yunnan Province, Kunming, 650091, China; slqin@bao.ac.cn}
\altaffiltext{2}{Physikalisches Institut, Universit\"at zu K\"oln, Z\"ulpicher Str. 77, D-50937 K\"oln, Germany}
\altaffiltext{3}{Department of Physics and Astronomy, University
of California, Los Angeles, CA 90095, USA}
\altaffiltext{4}{Department of Astronomy, Peking University, Beijing, 100871,
  China}
\altaffiltext{5}{Korea Astronomy and Space Science Institute 776, Daedeokdaero, Yuseong-gu, Daejeon, Republic of Korea 305-348}
\altaffiltext{6}{Department of Physics and Hebei Advanced Thin Film Laboratory, Hebei Normal University, Shijiazhuang 050024, China}





\begin{abstract}

 We report on the Submillimeter Array (SMA) observations of molecular
lines at 270 GHz toward W3(OH) and W3(H$_2$O) complex. Although
previous observations already resolved the W3(H$_2$O) into two or
three sub-components, the physical and chemical properties of the
two sources are not well constrained. Our SMA observations clearly
resolved W3(OH) and W3(H$_2$O) continuum cores. Taking the
advantage of the line fitting tool XCLASS, we identified  and
modeled a rich molecular spectrum in this complex, including multiple
CH$_3$CN and CH$_3$OH transitions in both cores.  HDO,
C$_2$H$_5$CN, O$^{13}$CS, and vibrationally excited lines of HCN,
CH$_3$CN, and CH$_3$OCHO  were only detected in  W3(H$_2$O). We
calculate  gas temperatures and column densities for both cores.
 The results  show that W3(H$_{2}$O) has  higher gas
temperatures, and larger  column densities than W3(OH) as previously observed,
suggesting physical and chemical differences between the two cores. We compare the molecular abundances in
  W3(H$_2$O) to those in the Sgr B2(N) hot core, the Orion~KL hot core and
the Orion Compact Ridge, and discuss the chemical origin of specific species.
 An east-west velocity gradient is seen in W3(H$_2$O), and the
extension is consistent with the bipolar outflow orientation
traced by water masers and radio jets. A north-south velocity
gradient across W3(OH) is also observed. However, with current
observations we can not assure if  the velocity gradients
are caused by rotation, outflow or radial velocity differences of the
sub-components in W3(OH).

\end{abstract}




%% what keyword punctuat ion is appropriate.

\keywords{ISM:abundances --- ISM: individual (W3(OH))
--- ISM: molecules --- radio lines: ISM
--- star: formation}









\section{INTRODUCTION}
How massive stars form is poorly understood yet, partly because powerful
radiation pressure from stars with masses above $\sim$8 M$_{\odot}$
should prevent the gas accretion on the protostars to form more massive
stars, according to the monolithic collapse model. Different theoretical scenarios
related to high-mass star formation have been proposed, i.e., monolithic
collapse (e.g., Jijina \& Adams 1996; McKee \& Tan 2003), competitive
accretion and coalescence (e.g., Bonnell et al. 1997, 1998), and how each
of them takes effects depends on the initial environments of
their parent clouds (Zinnecker \& Yorke 2007, Tan et al. 2014). Observations are essential to test scenarios of high-mass star
formation. Lower spatial resolution observations can not resolve
 detailed structure due to their large distances, nor can they explore the
 clustered environments, and small scale variations in high-mass star forming regions. Higher spatial resolution observations are necessary to characterize
 kinematics, physical and chemical conditions of high-mass star forming
 regions at small spatial scales.

The W3(OH) complex, located at 2.04 kpc (Hachisuka et al. 2006), is
one of the nearest   and well-studied high-mass star forming regions, and
harbors two objects, W3(OH) and W3(H$_2$O). Radio observations
suggested that  W3(OH) is an  UC H{\sc ii} region, ionized by young OB stars,
and
is rich in OH masers (Reid et
al. 1995; Wilner et al. 1999; Fish \& Sjouwerman 2007).
W3(H$_2$O), also known as W3(OH)-TW (Turner \& Welch 1984), locates
6$^{\prime\prime}$ east of W3(OH) and is rich in
H$_2$O maser and organic molecules (Wyrowski et al. 1999; Chen et
al. 2006; Zapata et al. 2011; Hern\'andez-Hern\'andez et al. 2014), and presents
 hot core properties (Kurtz et al. 2000).  Single dish observations have
shown overall inflow in the W3(OH) complex (Wu \& Evans 2003), and
 outflows as well as a possible disk were identified by Zapata et al. (2011). Subarcsec
resolution observations of continuum and CH$_3$CN lines showed a
high-mass protobinary system in W3(H$_2$O), with the two sub-cores having
different physical properties (Chen et al. 2006). However, physical
and chemical properties of W3(H$_2$O) and W3(OH) are still not
well characterized.

In this paper, we present results from Submillimeter Array
(SMA\footnote {The Submillimeter Array is a joint project
between the Smithsonian Astrophysical Observatory and the Academia
Sinica Institute of Astronomy and Astrophysics and is funded by
the Smithsonian Institution and the Academia Sinica.}) observations
towards the W3(OH) complex with moderate spatial resolution. Our goal is to study
the physical and chemical differences between W3(H$_2$O) and
W3(OH). The observations were tuned to 267 GHz covering  the linear molecule HCN
and other complex molecules. Compared to other submillmeter bands,
 molecular emissions at 267 GHz have less line confusion
(Greaves \& White 1991).  We describe the observations in \S 2.
 In \S 3 we present the spectral line results, followed by data analysis in \S
 4.  \S 5 discusses differences of physics and chemistry of the
 two sources.  We summarize the results in \S 6.

\section{OBSERVATIONS}

 The track-sharing SMA observations were carried out with seven antennas on 2007,
November 30, in its compact array, for a total of 10.23 hours on
calibrators and the target sources (W3(OH) and S231). The
phase-tracking center of the W3(OH) complex was at R.A.=$02^{\rm
h}27^{\rm m}04^{\rm s}.68$,
decl.=$+61^{\circ}52^{\prime}25^{\prime\prime}.5$ (J2000.0). The
database covers 4 GHz bandwidth, ranging from 265.65 to 267.65 GHz
(lower sideband), and from 275.65 to 277.65 GHz (upper sideband).
The frequency resolution of 0.406 MHz corresponds to a velocity
resolution of $\sim$0.5 km s$^{-1}$.  The average zenith
opacity ($\tau_{225 \rm GHz}$)  measured by the tipping
radiometer at the Caltech Submillimeter Observatory (CSO) was
0.12, indicating moderate weather conditions during observations.  Uranus
($\sim$ 45~Jy at 1.1~mm band) was observed for bandpass
calibration. The QSOs 0136+478 ($\sim$1.8 Jy) and 3c111 ($\sim$7.2
Jy) were observed in the ascending time order, for phase
correction.  Flux calibration is based on observations of
Uranus and a model of its brightness distribution. Comparing with
flux derived from the quasar monitoring, the flux calibration is
estimated to be accurate to within 20\%.

The data were calibrated and imaged in Miriad (Sault, Teuben \& Wright 1995). We selected line-free
channels for continuum-subtraction in the (u, v) domain using the UVLIN
task. The projected baselines ranged from 8 to 64~k$\lambda$. The
resulting synthesized beam is
$\sim$2$^{\prime\prime}.7$$\times$2$^{\prime\prime}.4$
(PA=--61$^{\circ}$). The continuum image was obtained by averaging all
line-free channels of the lower and upper sidebands resulting in a
sensitivity of 0.02~Jy~beam$^{-1}$. The 1 $\sigma$ noise level
of the spectral line images is 0.1 Jy~beam$^{-1}$ per channel. 1
Jy~beam$^{-1}$ in these observations corresponds to a main beam
brightness temperature of $\sim$2.4 K.

\section{RESULTS}
\subsection{\emph{Continuum}}
Figure 1 presents the synthesized 1.1 mm continuum image, in which
the two cores, W3(H$_2$O) and W3(OH), are well resolved. The
peak intensity, total flux density, and deconvolved source size
from a 2D Gaussian fit to the two continuum cores are summarized in Table 1.
 W3(H$_2$O) and W3(OH) have comparable peak intensities, but
different flux densities and source sizes. The dust mass and
source-averaged H$_2$ column density  of W3(H$_2$O) can be calculated by the
formulae (Hildebrand 1983)

\begin{equation}
M_{\rm gas}= \frac{S_{\nu}D{^2}R}{\kappa_{\nu}B_{\nu}(T)},
\end{equation}

\begin{equation}
N(H_{2})= \frac{S_{\nu}R}{2m_{H}\Omega \kappa_{\nu}B_{\nu}(T)},
\end{equation}

\noindent
where $S_{\nu}$ is the continuum flux, D is the distance to
source, R is the gas-to-dust ratio (100), $\Omega$ is the solid
angle  subtended by the source, and B$_{\nu}$(T) is the Planck function
at temperature T. The dust mass opacity coefficient $\kappa_{\nu}$
of 1.55 cm$^{2}$ g$^{-1}$ is interpolated from the values of
no-coagulated dust grains with thin ice mantles (Ossenkopf \&
Henning 1994). A dust temperature of 100~K is adopted for  the W3(H$_2$O) 
core (Chen et al. 2006; Zapata et al. 2011). The derived H$_2$
  column density and mass (listed in Table 1) are slightly
  larger than those derived by Hern\'andez-Hern\'andez et al. (2014), probably due to
  the lower angular resolution of our observations that allow us to recover more
  extended emission. The continuum source that we detect at the position of W3(OH) is mainly
  free-free emission from the H{\sc ii} region associated with it (Willner et al. 1995; Wyrowski et al. 1997, 1999). An
  upper limit of 0.5~Jy at 220~GHz is inferred from dust emission (Wyrowski et
  al. 1999). Therefore, our continuum observations can not provide a reliable estimation for the
  H$_2$ column density towards W3(OH).

\subsection{\emph{Molecular lines}}

We identify line transitions using XCLASS\footnote{http://www.astro.uni-koeln.de/projects/schilke/ XCLASS} package.
 XCLASS accesses the CDMS (M\"uller at al. 2001, 2005; http://www.cdms.de) and JPL (Pickett
et al. 1998; http://spec.jpl.nasa.gov) molecular databases which
provide all necessary spectroscopic information: rest frequency,
integrated intensity, lower state energy, upper state degeneracy,
quantum numbers and the partition function. Figure 2(a) and (b)
show the lower and upper sideband spectra extracted from   the
image domain in both cores W3(H$_2$O) and W3(OH), with the different
line identifications. At a first glance, there are more
lines in W3(H$_2$O) than in W3(OH). Strong emission of rotational
transitions of HCN, HCO$^{+}$, OCS, $^{13}$CS, CH$_3$OH,
CH$_3$OCHO and CH$_3$CN  as well as highly excited CH$_3$OH transitions are present in both cores. While
vibrationally excited lines HCN v$_2$=1, CH$_3$OCHO v=1,
CH$_3$CN v$_8$=1, as well as rotational transitions of
C$_2$H$_5$CN, HDO, CH$_2$NH, O$^{13}$CS are only detected in
W3(H$_2$O). The CH$_3$OH rotational transitions in
W3(OH) have higher intensities than in W3(H$_2$O),  while the
vibrational lines appear stronger in W3(H$_2$O).
Note that five CH$_3$$^{13}$CN transitions were detected only in
W3(H$_2$O) (see Figure 2(c) for an enlarged version).
 The spectral differences between W3(OH) and  W3(H$_2$O) are probably caused by
 excitation conditions or molecular abundances, in which  W3(H$_2$O) has hotter
gas environments and higher abundances than W3(OH) (Hern\'andez-Hern\'andez et al. 2014; also see discussion in sections
4.1 and 4.2).  The H$_{40}\gamma$ recombination line at 276.746 GHz is observed in W3(OH), but
  not in W3(H$_2$O), confirming the presence of a bright H{\sc ii} region in W3(OH).

\subsection{\emph{Gas distribution}}

  Previous higher spatial resolution
continuum observations already resolved W3(H$_2$O) into three
sub-components: A, B and C (Wyrowski et al. 1999). The line
observations suggest that nitrogen-bearing molecules only peak
at component C and oxygen-bearing molecules have a different
distribution (Wyrowski et al. 1999). Figure 3 presents maps of different molecular species.
C$_2$H$_5$CN and HCN vibrational transitions are only detected towards
 W3(H$_2$O). While CH$_3$CN, CH$_3$OCHO, and higher energy CH$_3$OH (at
 266.704 GHz) transitions are present in both cores, with higher intensities observed towards
W3(H$_2$O). The morphology of CH$_{3}$OH at 266.838 GHz is almost the same as the
continuum image (see Figure 1)  which may indicate its grain surface
origin. The line emission and gas distribution indicate different physical and chemical conditions
for the two sources.

\subsection{\emph{Velocity structure}}
The line velocities generally trace kinematics in molecular clouds.
Mean systematic velocities of $-$49.4 and $-$45.1~km~s$^{-1}$ for  W3(H$_2$O)  and W3(OH) are obtained from modeling
of various molecular lines (see Table 2 and next section), which is roughly consistent with the values
derived from CH$_3$CN lines by Hern\'andez-Hern\'andez et al. (2014). The systematic velocity variations among the different
species are also observed in W3(H$_2$O), which are likely
caused by the chemical differences of the molecules  at a small scale.
  High spatial resolution observations of CH$_3$CN lines resolve the A and C components in W3(H$_2$O) and reveal
different systematic velocities of  $-$51.4 and $-$48.6~km~s$^{-1}$ (Chen
et al. 2006).  Systematic velocities of  $-$50.5 and $-$46.5~km~s$^{-1}$
 for the A and C components are derived by Zapata et al. (2011). In our SMA
observations, HCN v$_2$=1 and C$_2$H$_5$CN appear to peak at the
same position, and have a systemic velocity  ($-$48.5
km~s$^{-1}$) similar to that of component C. While CH$_{3}$CN, CH$_{3}$OH and
CH$_{3}$OCHO peak at a different positions, and have a systemic
velocity  $-$49.6 km~s$^{-1}$, which is not consistent with that of
components A or C.
Future observations of various species  with higher spatial,
spectral resolution, and better sensitivity can resolve the
detailed structure, and make an overall  physical and chemical
picture of this high-mass star formation region.

  In order to display the detailed velocity structure and kinematics, we present velocity channel maps of HCN v$_2$=1 at
267.199 GHz and the lower energy level CH$_{3}$OH rotational
transition at 266.838 GHz in Figure 4. In addition to the compact
gas emission concentrated on the continuum peak around the systematic
velocity, the HCN v$_2$=1 blue-shifted emission with velocity
ranging from $-$54 to $-$50 km~s$^{-1}$ locates to the east of
 the W3(H$_2$O) continuum peak, while red-shifted emission from $-$46 to
$-$42 km~s$^{-1}$ is located west of the W3(H$_2$O) continuum peak. The
 east-west velocity gradient across W3(H$_2$O) is also seen in intensity-weighted velocity map
 (moment 1) of HCN v$_2$=1 line, as shown in the left panel of Figure 5.
There is no HCN v$_2$=1 emission towards W3(OH). The blue-shifted
feature in  W3(H$_2$O) is also seen in lower energy
CH$_{3}$OH transition. The velocity gradient seen in the east-west
direction was also reported using other molecular tracers (Wyrowski
et al. 1997; Chen et al. 2006), and is consistent with the outflow
orientation traced by water masers and radio jets (Alcolea et al.
1993; Wilner et al. 1995).  Additionally, a north-south velocity gradient
 in W3(OH) is observed in the CH$_{3}$OH channel maps of Figure 4, covering a
 velocity range from $-$48 to $-$41 km~s$^{-1}$. The first order moment image of
 CH$_{3}$OH is shown in the right panel Figure 5, and it reveals the systemic velocity difference between W3(H$_2$O) and
W3(OH). However, whether the velocity gradient is caused
by outflow, disk, radial velocity differences of multiple sources, or
binary rotation is still unclear. Higher spatial and
spectral resolution observations of  multiple lines are needed to explain the
velocity gradient.

\section{ANALYSIS}

\subsection{\emph{Rotation temperature and column density}}

The rotation temperature diagram (RTD) method based on multi-transition
observations is commonly used for determining physical
parameters, in which population distribution of all energy levels
of a specific species is described by a single excitation
temperature and column density (Goldsmith \& Langer 1999). RTD can be safely used in an
homogeneous cloud with simple spectral profiles,  optically thin emission, and a large spatial
extension that completely covers the observing beam. However
 perfect homogeneous clouds are not observed in interstellar space,
and complex density and temperature structures have been observed even in
clouds with a simple morphology. Under assumption of local
thermodynamical equilibrium (LTE), the XCLASS program solves for
the radiative transfer equation which takes source size, beam
filling factor, line profile, line blending, background
temperature, excitation, and opacity
 into account. The main difference between RTD and XCLASS fitting
is that XCLASS can deal with opacity, line blending, and multiple
velocity and spatial components. The detailed fitting  functions
and modeling procedures are described in the papers by Comito et al. (2005)
and Zernickel et al. (2012). The source sizes are derived by
 two dimensional Gaussian fitting to the individual
spatial components and converted to circular sizes. Then, LTE modeling is performed on
 the observed spectrum (see Figure 2). The systemic velocities, line widths, rotation temperatures and column densities
of observed species derived with XCLASS modeling are listed in Table 2.  The
results show that different molecules have different rotational temperatures
and column densities. The differences in the physical (distribution in the cloud) and chemical properties of the different molecules
explain why not a single temperature is observed for all the species (see review by van Dishoeck \& Blake 1998).

 Note that in Figure 2C, the predicted intensities for the lines with optical depths
larger than 1 are stronger than those in the observed spectra,
while the predicted intensities for the lines with optical depths
less than 1 are coincident with the observed ones. For the optically thick
transitions, one just sees the $\tau$=1 surface due to photon trapping, while
the XCLASS models sum up all the contributions from the cloud linearly.

Multiple transitions of CH$_{3}$CN and CH$_{3}$OH  spanning a wide
energy range (the upper level energies are  labelled in
Figure 2) are observed in both W3(H$_2$O) and W3(OH), which can
provide constraints on parameter estimation, especially for gas
temperatures.  In principle, the higher energy transitions will
sample the emission from hot cores since these transitions are
populated in conditions of high densities and temperatures, while
the lower energy transitions can be excited in both the warm and
the cool surrounding regions. Therefore two components with
different temperatures and column densities are needed to model
CH$_{3}$CN and CH$_{3}$OH in the line abundant region W3(H$_2$O).
We discuss individual molecules in the following.

CH$_{3}$CN is thought to be a good
 probe of kinetic temperature of high density gas (e.g., Remijan et al.
2004; Schilke et al. 1997). Seven CH$_{3}$CN rotational transitions (with E$_u$ of 105--363 K)
are detected in
W3(H$_2$O). The rotational transitions of
CH$_{3}$CN in  W3(H$_2$O) can not be fitted by a single gas
temperature, two component modeling with gas temperatures of 55 (cold)
and 200 K (warm),  and column densities of 0.15$\times$10$^{17}$ and
0.3$\times$10$^{17}$ cm$^{-2}$ are needed. Five CH$_{3}$$^{13}$CN
transitions were also observed in W3(H$_2$O), of which three are blended with CH$_{3}$CN. Using the same
temperatures as for CH$_{3}$CN, the derived column densities are
 0.002$\times$10$^{17}$ and 0.004$\times$10$^{17}$ cm$^{-2}$ for the warm and cold
components, respectively. Six CH$_{3}$CN v$_8$=1 vibrational lines (with
E$_u$ of 625--692 K) are detected towards W3(H$_2$O). The LTE modeling of the CH$_{3}$CN v$_8$=1 gives a
rotation temperature of 200 K and a column density of
0.3$\times$10$^{17}$ cm$^{-2}$. The rotational
transitions of CH$_{3}$CN are also detected in W3(OH), but have
lower intensities than in  W3(H$_2$O). No CH$_{3}$CN  v$_8$=1 or
CH$_{3}$$^{13}$CN are
detected in this core. The observed spectra can be fitted by a
single gas temperature of 95 K and column density of
0.016$\times$10$^{17}$ cm$^{-2}$, which are lower than in W3(H$_2$O).

Three C$_{2}$H$_{5}$CN spectral features were
identified toward  W3(H$_2$O), but not in W3(OH). A rotation temperature of 105
K and column density of 0.08$\times$10$^{17}$ cm$^{-2}$ were derived.

One rotational transition of HCN was
detected in both cores showing complicated kinematics (infall and outflow profiles that are discussed in
Qin et al. 2015). Two vibrationally excited HCN transitions
(with E$_u$ of 1050 K) have been only detected in W3(H$_2$O), giving
a rotation temperature of 360 K and column density of
0.25$\times$10$^{17}$ cm$^{-2}$.

CH$_{3}$OH has been widely
detected in various star-forming regions. Four rotational
transitions (with E$_u$ of 57--690 K) and one vibrational transition (with
E$_u$ of 710 K) were detected in both
cores,  but high energy transition in  W3(H$_2$O) have higher
intensities than in W3(OH), while  a higher intensity for the lower energy
transitions is observed in W3(OH). Similar to CH$_{3}$CN, a two temperature component model is needed, giving rotation
temperatures of 55 and 360 K, a column densities of
2.4$\times$10$^{17}$ and 5.5$\times$10$^{17}$ cm$^{-2}$ in
W3(H$_2$O). One temperature component fit to W3(OH) gives a gas temperature
of 155 K, and  a slightly lower column density compared with W3(H$_2$O).

Nine weak spectral features of CH$_{3}$OCHO
 (with E$_u$ of 351--365 K) were identified in its vibrational state, and one
rotational transition  (with E$_u$ of 160 K) was detected in
W3(H$_2$O). Only one rotational transition was detected in W3(OH).
The derived  rotation temperature and  column density are 105
K and 1.1$\times$$10^{17}$ cm$^{-2}$ toward  W3(H$_2$O).

The molecules with only one transition detected are HDO, CH$_2$NH,
O$^{13}$CS, OCS, SO$_2$, and $^{13}$CS. Note that HDO, CH$_2$NH
and O$^{13}$CS were only detected towards W3(H$_2$O), and CH$_2$NH was
marginally detected. They have similar morphologies to the other species detected in
 W3(H$_2$O) or W3(OH). Assuming excitation temperatures of 105 K and 95 K
for W3(H$_2$O) and W3(OH), we derived the column densities for these molecules
(see Table 2).

\subsection{\emph{Abundance}}

The column density of  a specific molecule is related to its
opacity. Chemical models
use the fractional abundance of a molecule relative to H$_{2}$. We
derived the fractional abundances of the observed molecules relative
to H$_{2}$ by $f_{\rm H_{2}}=N_{T}/N_{\rm H_{2}}$ (see Table 2),
where $N_{T}$ is the total column density of a specific
molecule and $N_{\rm H_{2}}$ is the H$_{2}$ column density derived from the
continuum.  Since no reliable H$_{2}$ column density is available for W3(OH), we
only calculate fractional abundances of molecules in W3(H$_2$O). They are compared to other
star forming regions in section 5.2. In general, molecules in W3(H$_2$O) with a small
source size have a high gas temperature and fractional abundance. The observations of multiple CH$_3$CN transitions toward a
sample of hot cores showed that the sources with high gas temperatures have
larger fractional abundances (Hern\'andez-Hern\'andez et al 2014). These results
will provide important constraints on chemical models.

\subsection{\emph{Isotopic ratio }}

Both CH$_{3}$CN and CH$_{3}$$^{13}$CN were observed in
W3(H$_2$O), which is  useful for demonstrating the line blending and
opacity effects in the modeling calculation. The close-up view
of the CH$_{3}$CN and CH$_{3}$$^{13}$CN modeling is shown in Figure
2(c), in which red and green curves are the synthetic spectra of
CH$_{3}$CN and CH$_{3}$$^{13}$CN, respectively, using both warm
(200 K) and cold (55 K) components, and the calculated opacity is
shown in lower panel. The high energy transitions of CH$_{3}$CN
tend to be optically thin, and opacities of CH$_{3}$$^{13}$CN
transitions are much lower than those of CH$_{3}$CN.  The two component
modeling results suggested that W3(H$_2$O) has an inner structure that is
unresolved in these observations. The rotation temperature of 200 K
for the warm component is consistent with that measured with HNCO lines
(Wyrowski et al. 1999).  A ratio for $^{12}$C/$^{13}$C of 75 is obtained
from the derived column density of CH$_{3}$CN and its
isotopologues (see Table 2), which is consistent with the value of 76$\pm$7 determined by
Henkel, Wilson \& Bieging (1982).

The peak optical depth can also be estimated from the observed  intensity ratio of
CH$_{3}$CN and CH$_{3}$$^{13}$CN as $\frac{T_{\rm mb}(\rm CH_{3}CN )}{T_{\rm
    mb}(\rm CH_{3}^{13}CN )} \sim \frac{1-e^{-\tau(\rm
    CH_{3}CN)}}{1-e^{-\tau(\rm CH_{3}CN )/R}}$, where R is $^{12}$C/$^{13}$C
ratio. The observed peak intensities of CH$_{3}$CN ($15_2-14_2$)
and CH$_{3}$$^{13}$CN ($15_2-14_2$) are 16.7 and 1.7 K,
respectively.  Taking  $^{12}$C/$^{13}$C to be 76, we derive an
optical depth of 7.7 for CH$_{3}$CN ($15_2-14_2$) which is roughly consistent
with that determined from the XCLASS modeling (See Figure 2(C)).

\section{DISCUSSION}

\subsection{\emph{Detection and non-detection of species}}

We now discuss the detection and non-detection of some species in the
W3(OH) complex.  High and low excitation lines of CH$_3$OH
have been detected in both  W3(H$_2$O) and W3(OH), but the transition
with the lowest energy level (at 266.838 GHz) in W3(OH) has  a larger extension and higher intensity than in W3(H$_2$O), indicating a
lower excitation condition in W3(OH). The most abundant species
in these observations is CH$_3$OH. Vibrationally excited
lines of HCN v$_2$=1, CH$_3$OCHO v=1,  CH$_3$CN v$_8$=1, and
rotational transitions of C$_2$H$_5$CN, HDO, CH$_2$NH, O$^{13}$CS
are only detected in W3(H$_2$O), and they have lower abundances
when compared to CH$_3$OH. Probably the lower abundances make these
transitions too weak in W3(OH), and the expected intensities are lower than
the SMA detection limit. We take O$^{13}$CS as an example to test this
assumption. In  W3(H$_2$O), the column density ratio of OCS and O$^{13}$CS is
80 (See Table 2). The column density of  OCS  in W3(OH) is
0.25$\times$10$^{17}$~cm$^{-2}$.  Assuming that OCS and O$^{13}$CS in W3(OH)
have  the same source size, gas temperature, and a column density ratio of 80,
  we simulated  O$^{13}$CS emission using XCLASS, and the predicted
  O$^{13}$CS intensity is much lower than the 1 $\sigma$ detection limit of our
  SMA observations. Therefore we argue that the
detection and non-detection of some species indeed reflect
physical and chemical differences between W3(H$_2$O) and W3(OH).

\subsection{\emph{Individual species}}

 Three regions: the Sgr B2(N) hot core, the Orion~KL hot core and Orion Compact
  Ridge, have the richest line emission in our Galaxy, and are good targets with which compare our results.
  In our observations towards W3(H$_2$O), the species CH$_{3}$OH,  CH$_{3}$CN and  CH$_{3}$OCHO, are detected in more than three transitions. In
  Figure 6, we show the abundances of the three species relative to H$_{2}$,
  compared to Herschel/HIFI observations of the Sgr B2(N) hot core (Neill et
  al. 2014), the Orion~KL hot core and Compact Ridge (Crockett et al. 2014).
  The comparison is reasonable, since the results of the Sgr B2(N)
  and Orion KL presented in Neill et al. (2014) and Crockett et al. (2014),
  and the results of W3(H$_2$O) presented in this work, have been obtained using the same tool, XCLASS. A possible source of
  uncertainty is the adopted H$_2$ column density. The abundances  among
  the four sources are similar, suggesting that these molecules have a same
  chemical origin in  W3(H$_2$O) as in the Orion KL and Sgr B2(N) sources,
  i.e., they originate from warm gas
  environments where some species are released from grain surfaces and involved into the
  gas phase chemistry. We discuss these molecules in the following.

The derived rotation temperatures for CH$_{3}$OH in W3(H$_2$O)
suggest that there are cold and hot gas components, and the higher
gas temperature component has a larger fractional abundance.
Laboratory works also confirmed that some complex species with
higher gas temperatures also have larger abundances (Fortman et
al. 2010, 2014). Compared with other organic species in these
observations, CH$_{3}$OH  has the highest fractional abundance.
Similar cases are also reported in other star forming regions
(e.g., Ge et al. 2014; van der Tak, van Dishoeck \& Caselli 2000;
Qin et al. 2010;  Crockett et al. 2014; Neill et al. 2014). Infrared observations suggest that CH$_{3}$OH
is the second most abundant ice species and has the highest abundance
relevant to water ice with a fraction of 5--30\% (e.g., Dartois et
al. 1999).  The high abundance of CH$_{3}$OH indicates that this
species may originate from grain surface chemistry (Charnely et
al. 2004; Garrod \& Herbst 2006).  The morphology of the lower energy
CH$_{3}$OH transition at 266.838 GHz is almost the same as the
continuum emission, which provides another support for the grain surface
origin of CH$_{3}$OH.   CH$_{3}$OCHO is detected in the Herschel/HIFI
survey of the Compact Ridge, but not in the Orion~KL hot core and Sgr B2(N) (Crockett et
al. 2014; Neill et al. 2014). This molecule may derive from CH$_{3}$OH on
grain surface (Garrod \& Herbst 2006) or in the gas phase (Laas et al. 2011).

 Both rotational and vibrational transitions of  CH$_3$CN were detected in W3(H$_2$O). The derived rotation
temperature of 200 K for the warm component is consistent with that obtained by
Wyrowski et al. (1999). The warm component has a higher fractional
abundance than the cold component,  which is consistent with
the trend that abundance increases with rotation temperature
derived by Hern\'andez-Hern\'andez et al. (2014). Also, a similar abundance
is measured in the Orion KL and Sgr B2(N). These facts
support that CH$_3$CN is synthesized by high
temperature gas phase reactions (Hern\'andez-Hern\'andez et al. 2014;
Neill et al. 2014).  There is no detection of CH$_{3}$CN vibrational transitions in
W3(OH). The intensities of rotational transitions of  CH$_{3}$CN in W3(OH)
are much lower than those  in  W3(H$_2$O), too. The derived  column density is one order of magnitude lower than that in
W3(H$_2$O).

The HDO transition at 266.161 GHz has been detected in the
OMC1 cloud (Greaves \& White 1991). HDO was only detected in
W3(H$_2$O) in our SMA observations. An HDO abundance relative to H$_2$ of
1.5$\times$10$^{-8}$ is derived. As stated above, CH$_{3}$OH may
originate from grain surface chemistry. If taking the abundance of
CH$_{3}$OH relative to water as 10\%, one obtains
HDO/H$_2$O=1.6$\times$10$^{-2}$, which may indicate that  W3(H$_2$O) is
at an early evolutionary stage (Miettinen, Hennnmann \& Linz 2011).

\subsection{Chemical difference between W3(H$_2$O) and W3(OH)}

 There are more lines in W3(H$_2$O) than in W3(OH). Furthermore, for the same
species, higher gas temperatures and column densities are obtained in W3(H$_2$O),
which indeed reflects different physics and chemistry between the two sources.
The differences may come from the fact that W3(H$_2$O) is at an early
evolutionary stage of high-mass star formation (e.g., Wyrowski et al. 1999;
Chen et al. 2006), while  W3(OH) is an expanding shell-like H {\sc ii} region
(Dreher \& Welch 1981; Kawamura \& Masson 1998).  The central star in W3(OH) is
optically obscured by a dusty cocoon (Wynn-Williams et al. 1972). Therefore
the gas seen in W3(OH) probably comes from the outer region of the hot core that
evolved into  the UC H{\sc ii} region, and the inner portion of this hot
core has been dissociated and ionized by  the stars.

\section{SUMMARY}

1. To characterize the physical and chemical differences between
 W3(H$_2$O) and W3(OH), we have carried out high spatial resolution
multi-line observations with the SMA at 270 GHz. The SMA
observations clearly resolved the two sources, W3(H$_2$O) and
W3(OH) continuum cores. More lines are detected in  W3(H$_2$O)
than in W3(OH). Rotational transitions of CH$_3$OH, CH$_3$OCHO and
CH$_3$CN are detected in both cores. While vibrationally excited
lines of HCN v$_2$=1, CH$_3$OCHO v=1, CH$_3$CN v$_8$=1, and
rotational transitions of C$_2$H$_5$CN, HDO are only detected in
 W3(H$_2$O). These features confirm that W3(H$_{2}$O) does present hot core properties.

2.  We have modeled the observed molecular lines using the
XCLASS software, under the assumption of LTE, and taking into account the source size, beam
filling factor, line profile, line blending, background temperature, and
excitation effects. Rotation temperatures, column densities and
fractional abundances are derived. Generally, the rotation temperatures and
 column densities are higher in  W3(H$_2$O) than in W3(OH). These
properties indeed reflect physical and chemical differences between the two
sources.  The differences are caused by the fact that W3(H$_2$O) is a hot
  core, while the gas seen in W3(OH) seems to come from the outer region of the hot
  cores located outside of the H {\sc ii} region.

3.   The abundances of CH$_{3}$OH, CH$_{3}$OCHO and CH$_{3}$CN in
 W3(H$_2$O) is similar with  those in the Sgr B2(N) hot core, the Orion~KL hot core
 and Compact Ridge, suggesting that these molecules have similar chemical origins.  Among the molecules detected, CH$_{3}$OH has the highest gas
abundance, and the morphology of its lowest energy transition is coincident with that of the continuum, suggesting that
it likely originates from grain surface
chemistry.  The abundance of CH$_{3}$CN increases as
temperature rises, and probably
CH$_{3}$CN was synthesized in high temperature gas phase reactions.
Non-detection and detection of some species such as HDO,
O$^{13}$CS, C$_{2}$H$_{5}$CN in the two cores may be related to their molecular
abundance or evolution. Vibrationally excited HCN and CH$_{3}$CN were detected
in W3(H$_2$O), but not in W3(OH); this is probably due to  hot core
environment of the W3(H$_2$O). The estimated HDO abundance relative to H$_2$O in W3(H$_2$O) is
$\sim$1.6$\times$10$^{-2}$, indicating an early
evolutionary stage.

4. Spectral images showed compact source structure centered at
 W3(H$_2$O) or W3(OH). The velocity channel maps of HCN v$_2$=1 and
CH$_3$OH show a east-west velocity gradient in  W3(H$_2$O) which is
consistent with the outflow orientation traced by water masers and
radio jets.  A north-south velocity gradient is seen in W3(OH).

\begin{acknowledgments}

 We thank the anonymous referee for his/her constructive comments
on the paper.  This work has been supported by the National Natural Science
Foundation of China under grant Nos. 11373026, 11373009, 11433004,
11433008, U1331116, and the National Basic Research Program of
China (973 Program) under grant No. 2012CB821800, by Top Talents Program of
Yunnan Province and Midwest
universities comprehensive strength promotion project (XT412001,
Yunnan university), by the Deutsche Forschungsgemeinschaft, DFG
through project number SFB956.

\end{acknowledgments}

\clearpage

\begin{deluxetable} {lccccccc}
\tabletypesize{\scriptsize}
\tablenum{1}
\tablewidth{0pt}
\tablecaption{Physical Parameters Of the Continuum Sources}
\tablehead{\colhead{Name }& \colhead{$\Delta$R.A.} &
\colhead{$\Delta$decl} &
\colhead{$I_{\rm peak}$} &\colhead{$S_{\nu}$} &\colhead{Deconvolved Angular
  Size} &\colhead{$N_{\rm H_{2}}$}& \colhead{Mass} \\
&\colhead{$^{\prime\prime}$} & \colhead{$^{\prime\prime}$} & \colhead{
  Jy~beam$^{-1}$}&\colhead{ Jy}&\colhead{$^{\prime\prime}$}&
\colhead{cm$^{-2}$}&\colhead{(M$_{\odot}$)}}
\startdata
  W3(H$_2$O)&--0.5&--0.8&2.25$\pm$0.14&4.6$\pm$0.3&3.4$\times$1.9
  (-72.5$^{\circ}$)&2.6$\times10^{24}$&26 \\
 W3(OH)&--5.9&--0.9&2.28$\pm$0.18&3.8$\pm$0.3&2.5$\times$1.6 (73.4$^{\circ}$)&\dots&\dots \\
\enddata
\tablenotetext{\it }{ Note: The offset positions are relative to phase track center
  of the observations at R.A.=$02^{\rm h}27^{\rm m}04^{\rm s}.68$,
decl.=$+61^{\circ}52^{\prime}25^{\prime\prime}.5$ (J2000.0).}

\end{deluxetable}

\begin{deluxetable} {lcccccc}

\tabletypesize{\scriptsize}

\tablenum{2}

\tablewidth{0pt}

\tablecaption{The Parameters Derived From Molecular Lines}
\tablehead{ \colhead{Molecule }& \colhead{$\Theta$} &\colhead{$T_{\rm rot}$} &
\colhead{$N_{T}$} &
\colhead{$f_{\rm H_{2}}$}   &\colhead{$V_{LSR}$}& \colhead{$\Delta V$}\\
&\colhead{${\prime\prime}$}   &\colhead{(K)}&\colhead{(cm$^{-2}$)}&&
\colhead{(km~s$^{-1}$)}&\colhead{(km~s$^{-1}$)}}

\startdata

&&&W3(H$_2$O)&&&\\

\hline

 CH$_{3}$OH&

    2.86&55&2.4$\times10^{17}$&9.2$\times10^{-8}$&--49.6&7\\
    &0.9&360&5.5$\times10^{17}$&2.1$\times10^{-7}$&--50.6&4\\

CH$_{3}$OCHO&

     2.33&105&1.1$\times10^{17}$&4.2$\times10^{-8}$&--49.6&6\\

 HDO&

   1.43& 105&0.4$\times10^{17}$&1.5$\times10^{-8}$&--49.4&6\\

OCS &

    1.93&105&3.2$\times10^{17}$&1.2$\times10^{-7}$&--49.2&5\\

O$^{13}$CS &

    1.93&105&0.04$\times10^{17}$&1.5$\times$$10^{-9}$&--49.1&5\\

HCN v$_2$=1 &
   1.46&360&0.25$\times10^{17}$&9.6$\times10^{-9}$&--48.5&8\\

CH$_{3}$CN&

   1.37&55&0.15$\times10^{17}$&5.8$\times10^{-9}$&--47.2&6\\
  &0.9&200&0.3$\times10^{17}$&1.2$\times10^{-8}$&--49.7&4\\

CH$_{3}$$^{13}$CN&

   1.37&55&0.002$\times10^{17}$&7.7$\times10^{-11}$&--47.2&4.5\\
  &0.9&200&0.004$\times10^{17}$&1.5$\times10^{-10}$&--49.7&3\\

CH$_{3}$CN v$_8$=1&
   0.9& 200&0.3$\times10^{17}$&1.2$\times10^{-8}$&--49.7&4\\

C$_{2}$H$_{5}$CN&

  1.93&105&0.08$\times10^{17}$&3$\times10^{-9}$&--48&7\\

$^{13}$CS &
   2.85&105&0.014$\times10^{17}$&5.4$\times$$10^{-10}$&--49.4&4\\

SO$_2$&
    1.93&105&0.7$\times10^{17}$&2.7$\times10^{-8}$&--49.3&6\\

\hline
&&&W3(OH)&&&\\
\hline

CH$_{3}$CN&

    2.2& 95&0.016$\times10^{17}$&\dots&--46&5\\

CH$_{3}$OH&

   3.5&155&2.2$\times10^{17}$&\dots&--44.8&5\\

CH$_{3}$OCHO&

 3.8&95&0.4$\times10^{17}$&\dots&--44.7&5\\

OCS &

  2.7&95&0.25$\times10^{17}$&\dots&--44.3&5\\

$^{13}$CS &
  3.76&95&0.005$\times$$10^{17}$&\dots&--45.6&4.3\\

SO$_2$&
  2.1&95&0.15$\times10^{17}$&\dots&--45.2&5.6\\

\enddata

\end{deluxetable}

%% If you use the table environment, please indicate horizontal rules using


%% Do not put multiple tabular environments within a single table.

%% The optional \label should appear inside the \caption command.

\begin{figure}
\vspace{-0.1cm} \epsscale{1.} \plotone{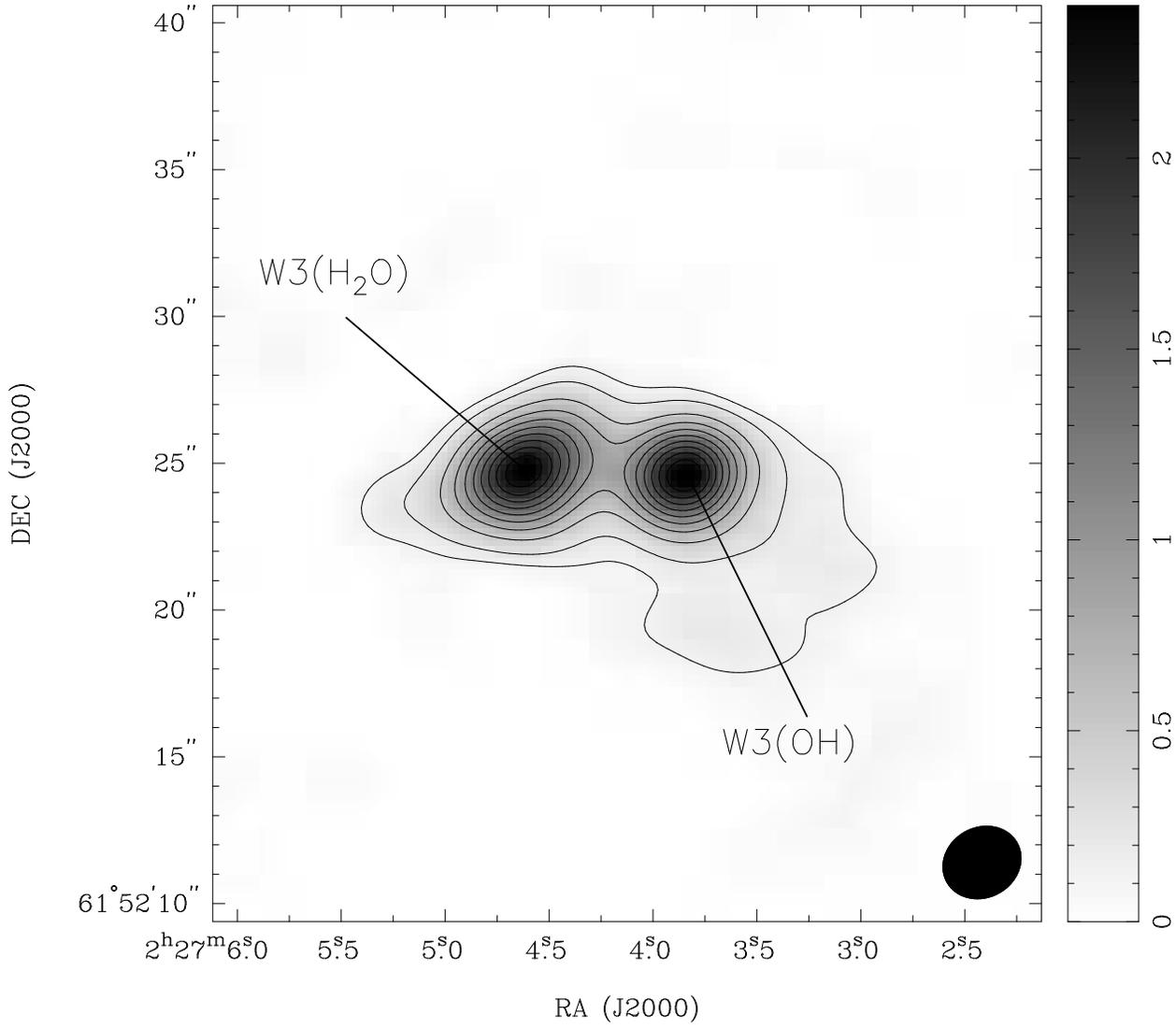}\vspace{-5.1cm}
\caption{Continuum image of W3(OH) complex in both contours and grey
  scale. The contours are from 10 to 90\% of the maximum intensity  (2.3
    Jy~beam$^{-1}$). The synthesized beam of
$\sim$2$^{\prime\prime}.7$$\times$2$^{\prime\prime}.4$
(PA=--61$^{\circ}$) is shown in lower right corner. }

\end{figure}

\begin{figure}
\vspace{-8mm} \epsscale{0.85} \plotone{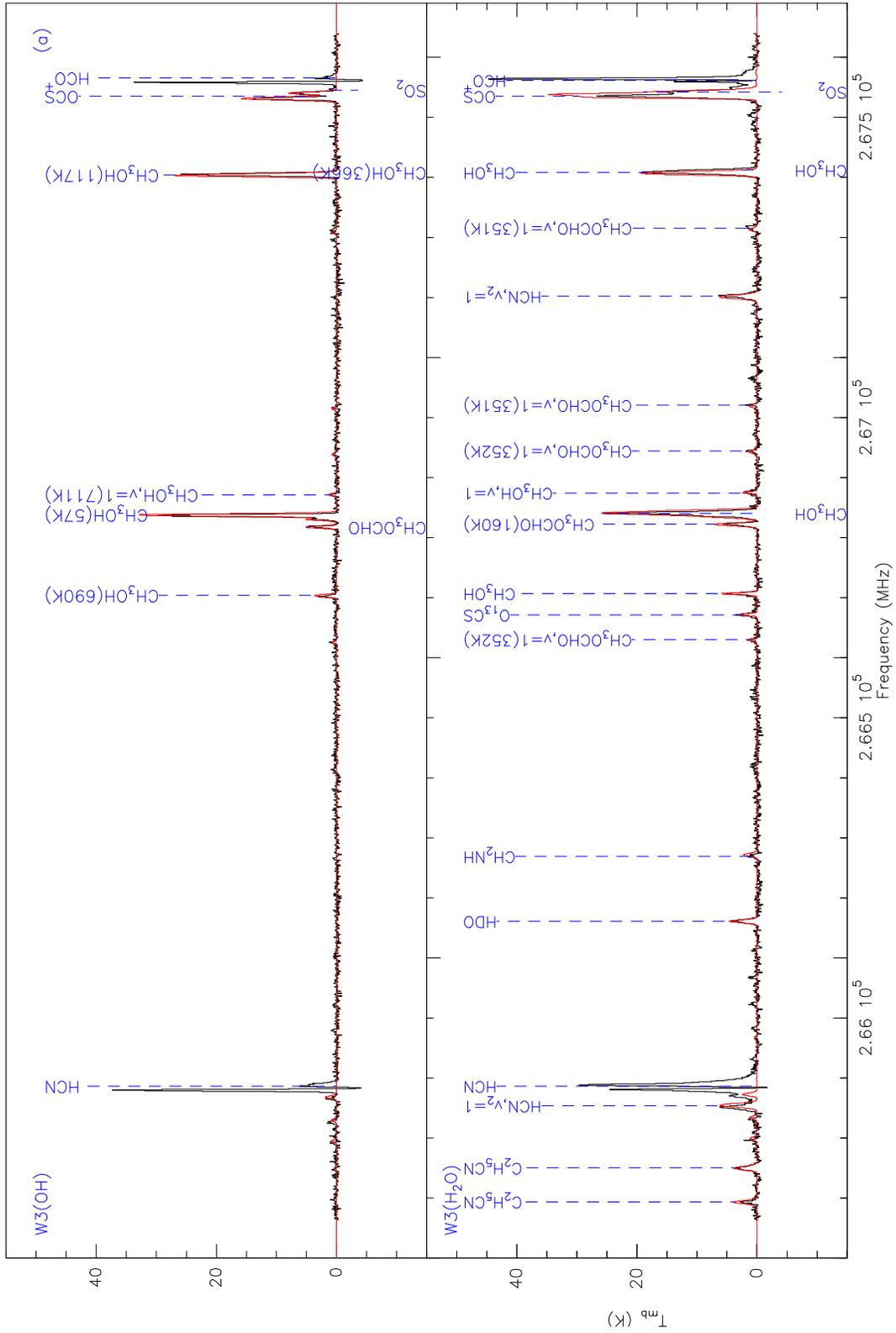}\vspace{-0.1cm}
\caption{A: Lower sideband spectra at  W3(H$_2$O) and
  W3(OH) continuum peak positions. The black curve is the observed spectral
  profile and red curve is from LTE modeling using XCLASS software. Upper
  panel is for W3(OH) and lower panel is for  W3(H$_2$O).  The upper level
  energies (E$_u$) of CH$_3$OH and CH$_3$OCHO are labelled. CH$_3$OH
(9$_{0,9}$-8$_{1,7}$) and (17$_{1,17}$-16$_{2,14}$) at 267.4 GHz are blended together.}

\end{figure}

\addtocounter{figure}{-1}

\begin{figure}

\vspace{-8mm}\epsscale{0.85}

\plotone{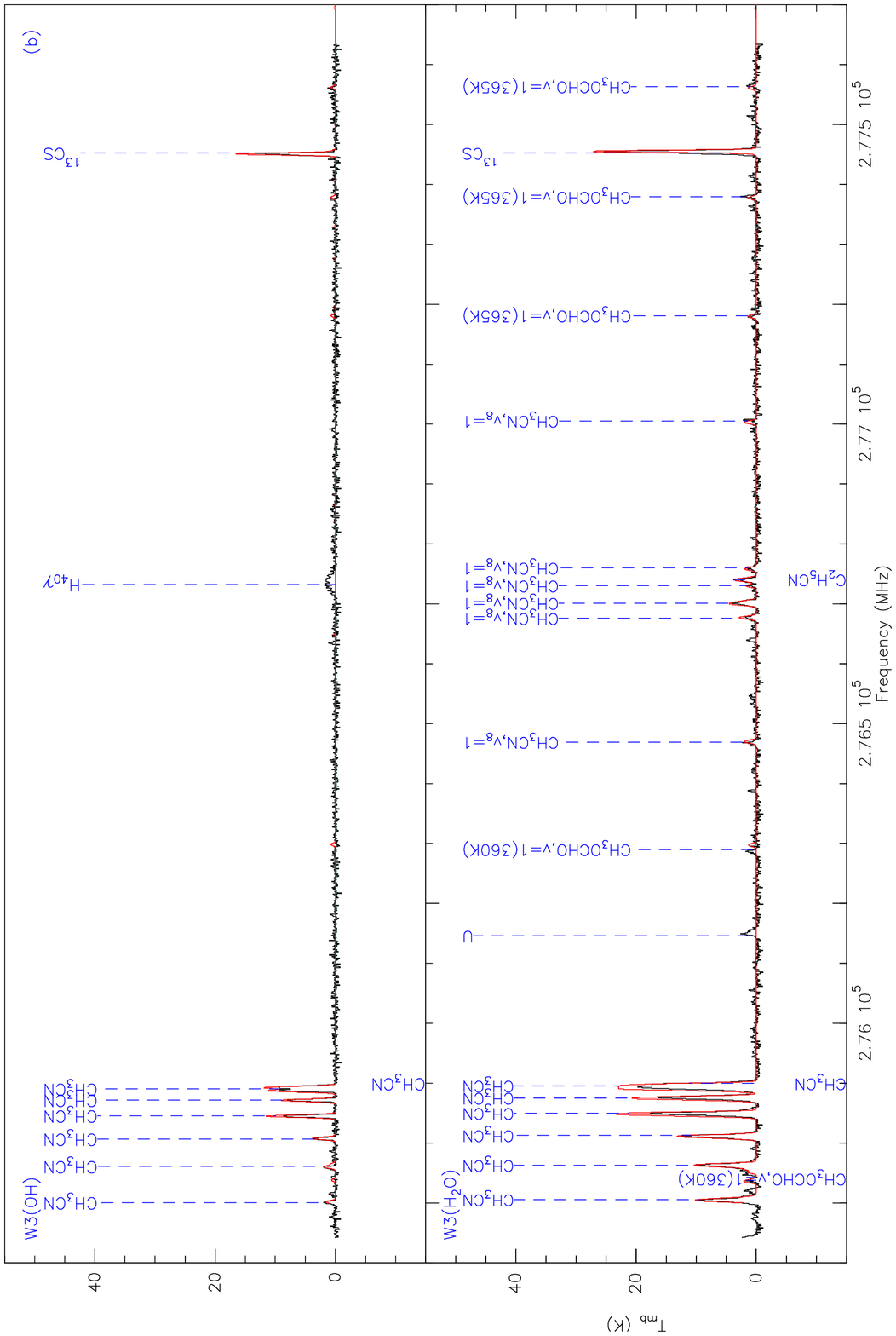}\vspace{-0.1cm}\ \caption{B: upper sideband spectra
  (CH$_3$$^{13}$CN is not labelled in this figure)  and XCLASS fitting. CH$_3$CN
(15$_{0}$-14$_{2}$) and (15$_{1}$-14$_{1}$) at $\sim$ 275.9 GHz are blended
  together.   The upper level
  energy (E$_u$) of  CH$_3$OCHO are  labelled. U indicates the unidentified lines.}

\end{figure}
\addtocounter{figure}{-1}

\begin{figure}

\vspace{-8mm}\epsscale{0.85}

\plotone{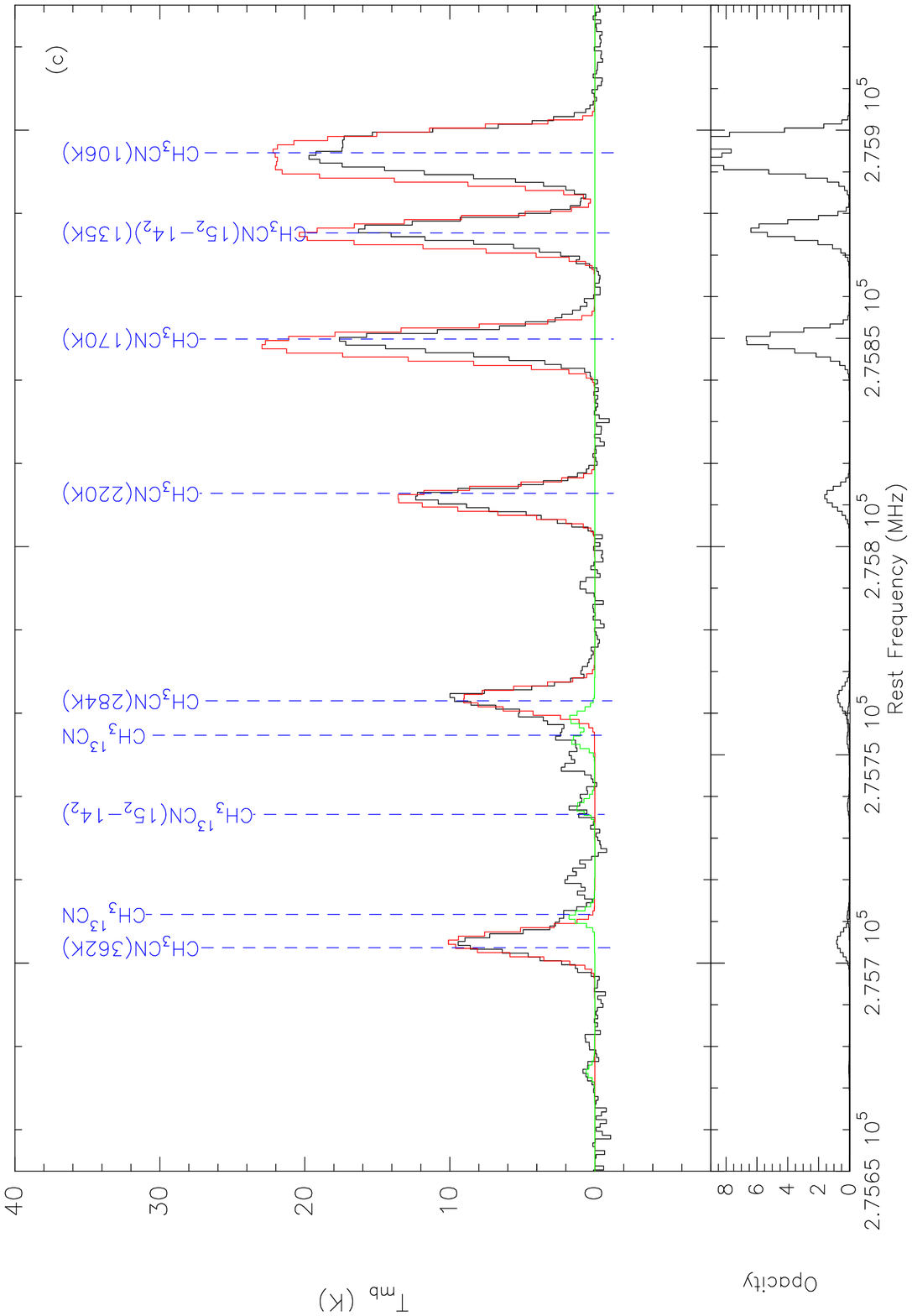}\vspace{-0.1cm}\ \caption{C: CH$_3$CN spectra and XCLASS fitting. In upper panel, the black curve indicates the observed
 spectrum, and red and green are the synthesized spectra for CH$_3$CN and CH$_3$$^{13}$CN), by use of both cold and warm components. In lower panel,
the black curve is the calculated opacity.  The upper level
  energy (E$_u$) of  CH$_3$CN are  labelled. CH$_3$CN (15$_{0}$-14$_{0}$) is
blended with (15$_{1}$-14$_{1}$) at 275.91 GHz; CH$_3$$^{13}$CN
(15$_{0}$-14$_{0}$) is blended with (15$_{1}$-14$_{1}$) at 275.77 GHz.}

\end{figure}

\begin{figure}
\epsscale{0.6}
\vspace{6mm}
\plotone{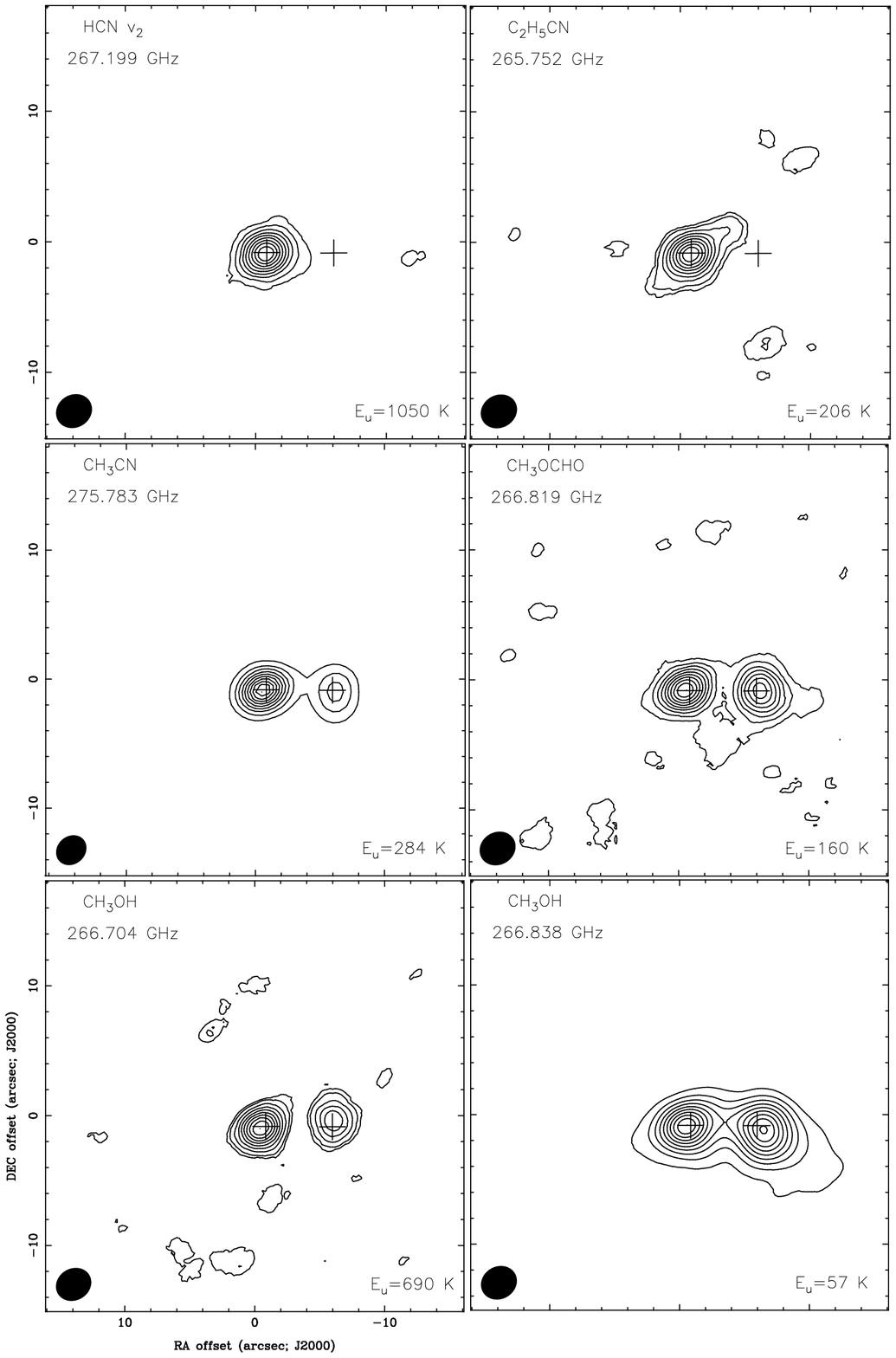}
\vspace{21mm}
\caption{Sample images from specific molecular species. The contours are from
  10 to 90\% of the maximum integrated intensity for each molecular
  transition.  The peak values of HCN v$_2$, C$_2$H$_5$CN, CH$_3$CN,
  CH$_3$OCHO, higher and lower energy transitions of CH$_3$OH are
  19.6, 11.6, 41.7, 12.2, 11.5, 78.5 Jy beam$^{-1}$ km s$^{-1}$,
  respectively. In each panel, the synthesized beam is shown in lower-right
  corner, and the cross symbol indicates the peak position of the continuum source. }

\end{figure}

\begin{figure}
\vspace{-3cm}\hspace{-5.2cm} \epsscale{1.3}
\plotone{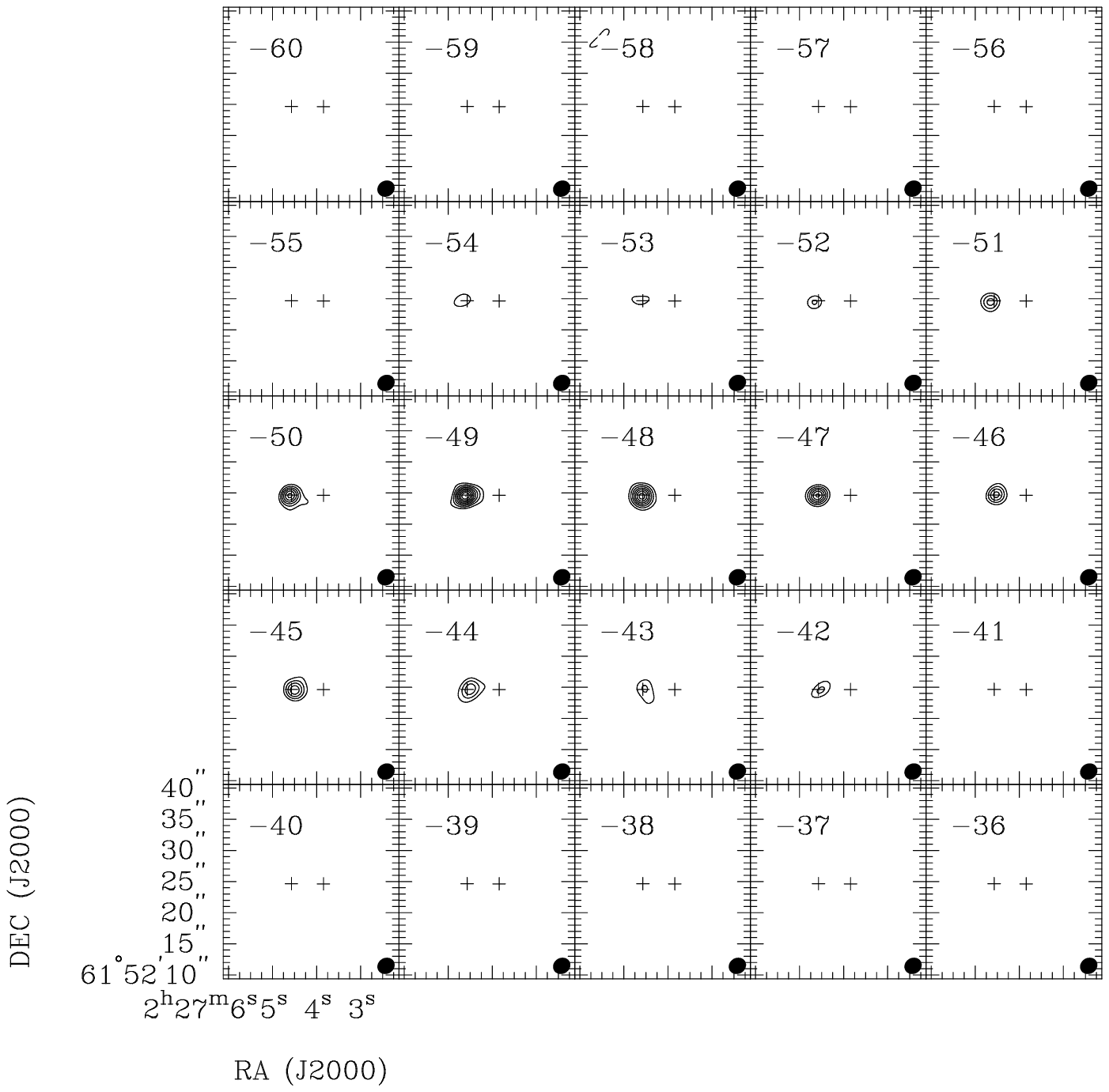}\vspace{-0.1cm}
\caption{  Channel maps of HCN v$_2$=1 at 267.199 GHz with FWHM beam size of
  $\sim$2$^{\prime\prime}.7$$\times$2$^{\prime\prime}.4$, PA=--61$^{\circ}$
  (lower right corner).  The contours are from
  10 to 90\% of the peak intensity (2.3 Jy beam$^{-1}$).  The rms (1 $\sigma$)
  noise level is 0.09 Jy beam$^{-1}$. In each
panel,  the cross symbol indicates
the peak position of the continuum source.  }

\end{figure}

\addtocounter{figure}{-1}

\begin{figure}

\vspace{-3mm}\hspace{-5.2cm}\epsscale{1.3}

\plotone{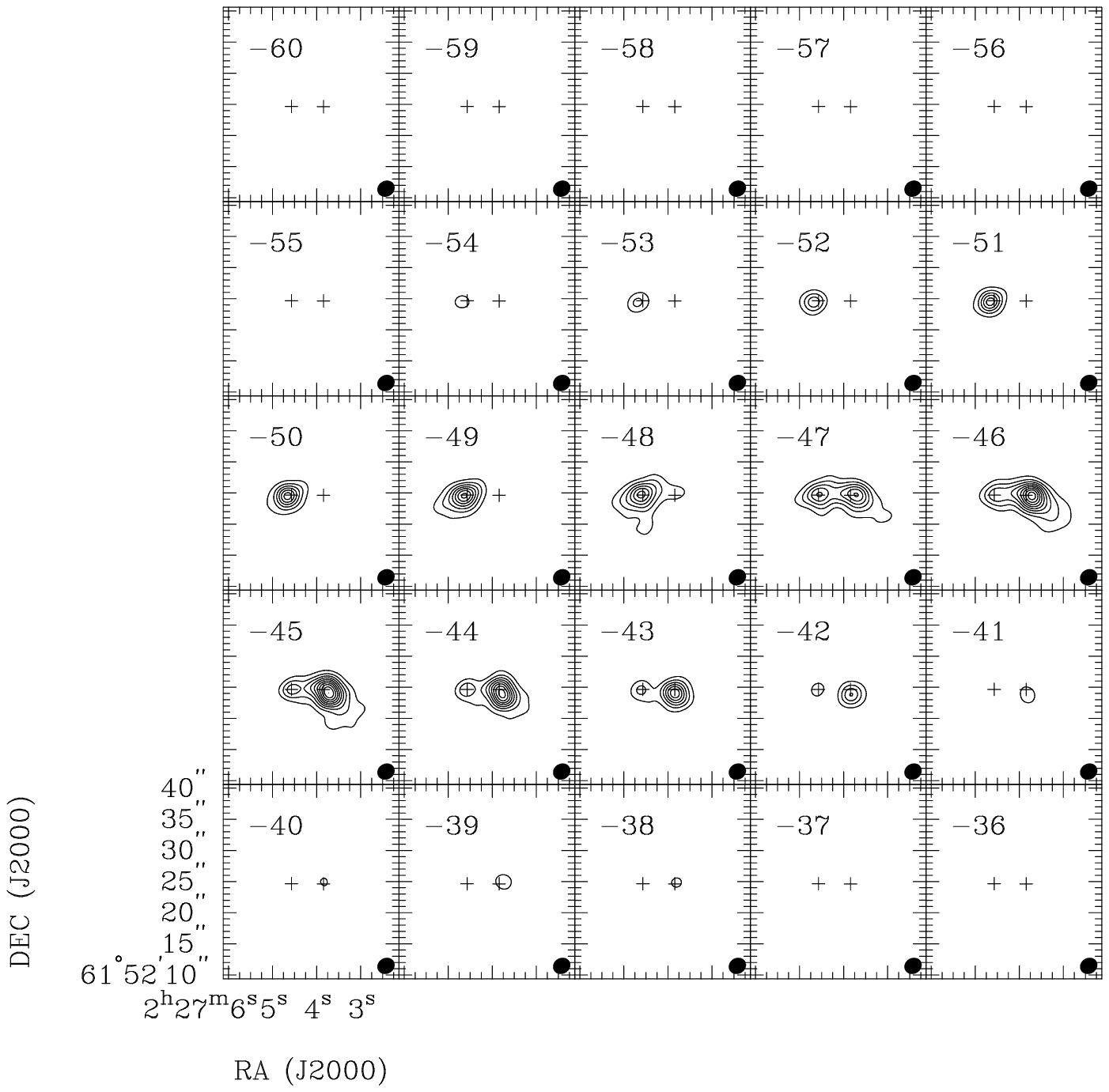}\vspace{-0.1cm} \caption{Continue:  Channel maps of CH$_3$OH at 266.838 GHz with FWHM beam size of
  $\sim$2$^{\prime\prime}.7$$\times$2$^{\prime\prime}.4$, PA=--61$^{\circ}$
  (lower right corner).  The contours are from
  10 to 90\% of the peak intensity (14.2 Jy beam$^{-1}$).  The rms (1 $\sigma$)
  noise level is 0.11 Jy beam$^{-1}$. In each
panel,  the cross symbol indicates
the peak position of the continuum source.}

\end{figure}

\begin{figure}

\vspace{-8mm}\epsscale{0.85}

\plotone{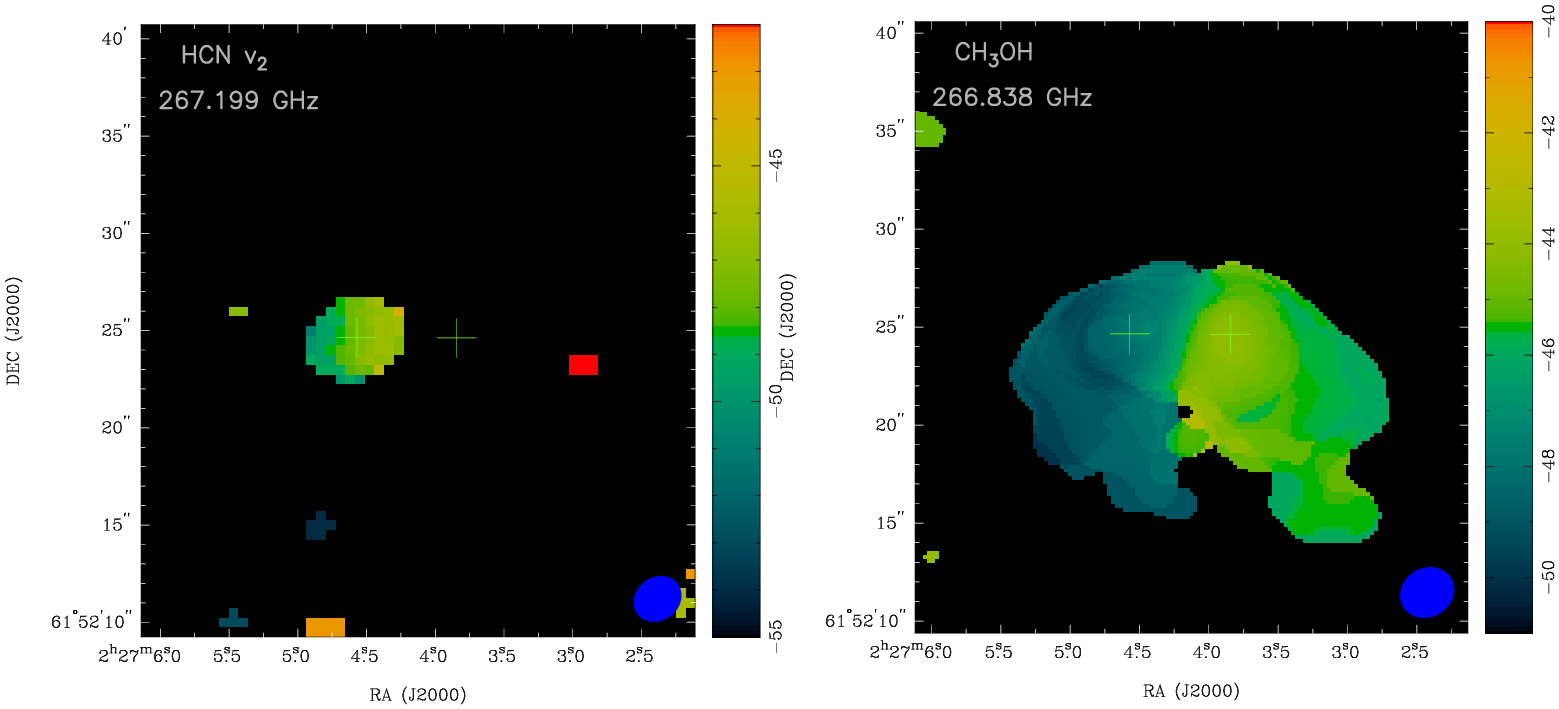}\vspace{-.1cm} \hspace{17.1cm}  \caption{Intensity-weighted velocity
  maps of HCN v$_2$=1 at 267.199 GHz, and   CH$_3$OH at 266.838 GHz with FWHM beam size of
  $\sim$2$^{\prime\prime}.7$$\times$2$^{\prime\prime}.4$,
  PA=--61$^{\circ}$. The scales on the right show the velocity range. }

\end{figure}
\begin{figure}

\vspace{-8mm}\epsscale{0.85}

\plotone{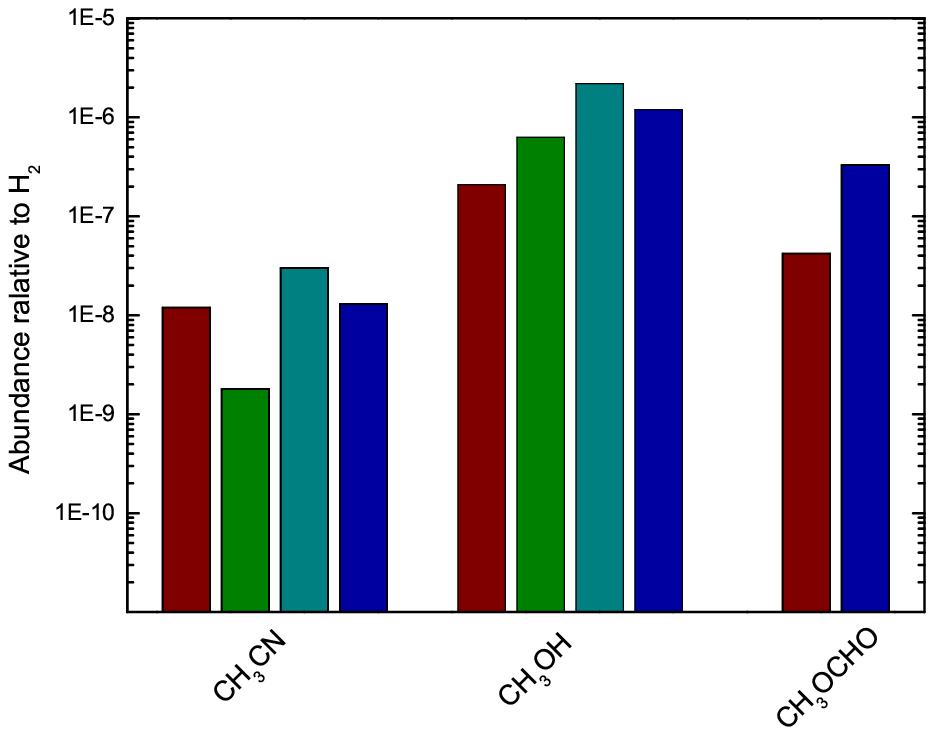}\vspace{-.1cm} \hspace{17.1cm}  \caption{Abundances of
  molecules in W3(H$_2$O) relative
  to H$_2$ (maroon bars), compared to abundances in the Sgr B2(N) hot core
  (green) from Neill et al. (2014), the Orion KL Hot core (turquoise)
  and Compact Ridge (blue) from Crockett et al. (2014).  }

\end{figure}
\end{document}